# Integrating Numerical Computation into the Modeling Instruction Curriculum

*Marcos D. Caballero,* Michigan State University, East Lansing, MI
*John B. Burk,* St. Andrew's School, Middleton, DE
*John M. Aiken and Brian D. Thoms,* Georgia State University, Atlanta, GA
*Scott S. Douglas, Erin M. Scanlon, and Michael F. Schatz,* Georgia Institute of Technology, Atlanta, GA

Numerical computation (the use of a computer to solve, simulate, or visualize a physical problem) has fundamentally changed the way scientific research is done. Systems that are too difficult to solve in closed form are probed using computation. Experiments that are impossible to perform in the laboratory are studied numerically. Consequently, in modern science and engineering, computation is widely considered to be as important as theory and experiment.

Unfortunately, most high school students today are never introduced to computation's problem-solving powers. Computer *usage* is widespread in high school STEM courses (e.g., obtaining lab data using computer acquisition hardware/software), but such usage rarely involves students constructing a computational representation of a science problem. The lack of computation in domain-specific STEM courses is not addressed in most high school computer science courses, which typically focus on programming and procedural abstractions rather than solving science problems. In recognition of these shortcomings, the recently published National Research Council's (NRC) framework for next-generation K-12 science standards lists "computational thinking" as one of the fundamental "practices" that should be incorporated into future K-12 science curricula.[1] The framework acknowledges that experience with computational thinking is crucially important, not only for developing future scientists and engineers, but also for providing all citizens with general insight into the science behind proposed solutions to technically complex social problems.

In this article, we describe a way to introduce physics high school students with no background in programming to computational problem-solving experiences. Our approach builds on the Modeling Instruction curriculum, which is currently used in approximately 10% of U.S. high school physics classrooms.[2] The Modeling Instruction approach emphasizes the practice of "Developing and using models" highlighted by the NRC K-12 science standards framework.[1] Coupling computational experiences with Modeling Instruction enables the modeling practice and the computational thinking practice to reinforce each other. To achieve this synergy, we taught 9th-grade students to use the VPython programming environment[3,4] within a Modeling-Instruction-based physics course.[5] We found that numerical computation within the Modeling Instruction curriculum provides coherence among the different models within the curriculum, links the various representations that the curriculum employs, and extends the curriculum to include real-world problems that are otherwise inaccessible using a purely analytic approach.

## Modeling Instruction & numerical computation

The Modeling Instruction curriculum employs a coherent framework of scientifically testing the limits of physical models (i.e., "the modeling cycle") by engaging students in the construction and comparison of different representations of physical phenomena.[6,7] Each modeling cycle is built on a set of modules; these modules promote scientific thinking through observation, experimentation, and discourse. By observing physical phenomena, representing those phenomena in a variety of ways, and making predictions of similar but not-yet-observed phenomena, students construct a working model that is able to fully describe the phenomena they observe. A full description of the modeling cycle is available in Refs. 6 and 7. Because of its emphasis on models, its focus on inquiry, and its use of multiple representations, the Modeling Instruction curriculum is effective not only in teaching students physical concepts,[5] but also in encouraging participation in class,[8] in helping align students' views about the nature of science with expert views,[9] and in promoting students' self-efficacy.[10]

Modeling Instruction treats each force and motion model as distinct, but the common thread of predicting motion using Newton's 2nd law and kinematics unifies them. The computational algorithm used to predict motion likewise retains the distinctions between the force and motion models, but highlights the commonality among them: namely, that such models differ only in the net force exerted on the system and in their particular initial conditions.

Given knowledge of the system's initial position and velocity, as well as the net force on the system, the algorithm for predicting motion can be described as a set of rules applied locally in space and time: (1) At a given instant in time $t$, compute the net force, $\mathbf{F}_{net}$, acting on the system, (2) For a short time $\Delta t$ later, compute the new velocity of the system using Newton's 2nd law, (3) At the same new time ($t + \Delta t$), compute the new position of the object using this updated velocity, and (4) Repeat Steps (1)-(3) starting at the updated time $t + \Delta t$. Formally, the iterative application of Steps (1)-(3) is, in effect, explicit (Euler-Cromer) numerical integration[11] of the equations of motion for Newtonian mechanics ($\Delta \mathbf{v} = \mathbf{a}\,\Delta t = \mathbf{F}_{net}/m\,\Delta t$, $\Delta \mathbf{x} = \mathbf{v}\,\Delta t$).

The mathematics behind iteratively predicting motion in this manner is well within the capabilities of most high school physics students (in either algebra-based or calculus-based courses); arguably, it is more accessible mathematically to students than the analytic methods currently used, even for the simplest cases (e.g., constant acceleration motion). Iterative motion prediction is usually too labor-intensive to perform by hand, but a computer can easily handle these calculations. Moreover, this same computational algorithm can be used simulate the vast majority of physical systems at a high-school level, further reducing the barrier for introductory students to explore complex systems.

Numerical computation offers significant pedagogical advantages. Computation highlights the relationship between the different physics models in the Modeling Instruction program (e.g., the no-forces model, the balanced-forces model, and the unbalanced-forces model). To produce simulations with qualitatively different behavior, we simply change the initial conditions (e.g., from 1D to 2D motion) or the net force (i.e., from constant to constantly changing). For example, we can generalize the balanced forces model to the unbalanced forces model by inserting a constant net force into the computational model. Furthermore, we can extend the unbalanced forces model to parabolic motion model by giving the object an initial velocity in both $x$ and $y$ directions.

Numerical computation provides dynamic animation and visualization of representations that are otherwise static in the Modeling Instruction curriculum. The output of numerical computation is continuously updating graphs (analogous to a chart recorder) and animations, not just numbers. The visualization provided by a numerical model is of paramount importance; certain aspects of visualization help students communicate a more coherent picture of their understanding.[12] These graphical and diagrammatic descriptions of the physical model, which might otherwise form the sole basis of the students' exposure to the model, are reproduced precisely by the computational model. Furthermore, the linking of representations can be done quite easily with a few simple lines of code (see next section, "Developing a set of computational tools").

These numerical models are not limited to analytically tractable solutions. This allows students to explore their real-world, rather than laboratory-constructed observations. Numerical computation provides a platform to focus class discussion on modeling and investigation without the undue burden of sophisticated mathematical techniques. For example, students observe objects that experience drag in their daily lives (try kicking a soccer ball!), and yet a model of this phenomenon is not explored in most introductory physics courses. A model of turbulent drag is a simple model to construct and describe. We have found that students can construct a model for drag, make sense of the model's predictions, and compare those predictions to those of the constant acceleration model (see section below, "A Typical activity: Modeling a kicked soccer ball").

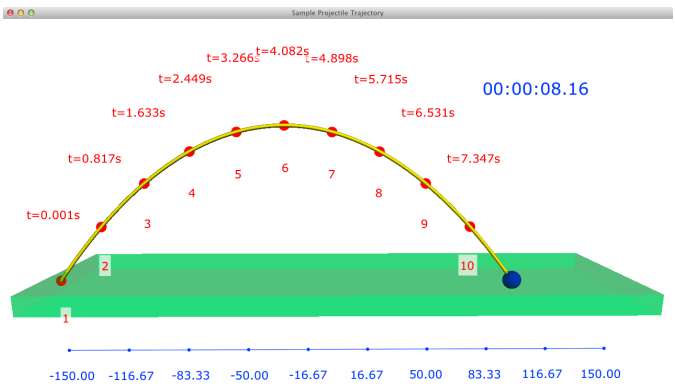

Fig. 1 - The visual output of a VPython+PhysUtil model of a soccer ball kicked in the air (without drag) constructed by three 9th grade students. PhysTimer appears in the upper-right corner (blue text). PhysAxis appears under the ground (blue line and text). MotionMap generated the ``breadcrumbs'' for the motion with time stamps and integer ordering (red spheres and red text).

By learning to use numerical computation, students acquire a familiarity with the tools of modern scientists and engineers. Moreover, as students gain experience with numerical computation, they begin to build "computer models" as part of their normal practice of constructing and testing models, which further emphasizes that models are what scientists and engineers use to describe physical phenomena. Numerical computation can be an effective tool for exploring the limits and refining the physical model in question. Students can explore the influence of such inputs on the resulting motion by changing parameters in the computational model. They can engage in prediction and confirmation by reviewing the animation and graphical output that their computational model produces.

## Developing a Set of Computational Tools

Numerical computation can provide additional benefits to student understanding of science and, in particular, physical phenomena. However, the tools that students use to numerically model such phenomena must include no more programming than is necessary. Their physics class is not a computer science course; hence, the program statements that students write should only reflect the representations with which they are becoming familiar. For our implementation, we used the VPython programming environment and employed it to focus students' computational model development on the physics of the particular system and the representations of that model. Moreover, we have developed a module,

```
1   from __future__ import division
2   from visual import *
3   from physutil import *
4   from visual.graph import *
5
6   track = box(pos=vector(0, -0.05, 0), size=(5.0, 0.05, 0.10),
              color=color.white)
7   cart = box(pos=vector(-track.length/2, 0, 0), size=(0.1, 0.04, 0.06),
              color=color.green)
8
9   mcart = 0.80                                    Initial Conditions
10  vcart = vector(3, 0, 0)
11
12  deltat = 0.01
13  t = 0
14  tf = 6.45
15
16  timerDisplay = PhysTimer(1, 1)
17  graph = PhysGraph()
18  axis = PhysAxis(track, 10, axisColor=color.red)
19  motionMap = MotionMap(cart, tf, 10, markerType="breadcrumbs",
                    labelMarkerOffset=vector(0,.3,0), dropTime=True)
20
21  while t < 2.3:
22
23      Ffan = vector(-0.75, 0, 0)
24      accel = Ffan/mcart                          Newton's Second Law
25      vcart = vcart + accel * deltat              + Kinematics
26      cart.pos = cart.pos + vcart * deltat
27
28      motionMap.update(t)
29
30      t = t + deltat
31      timerDisplay.update(t)
32
33      graph.plot(t, cart.pos.x)
```

Fig. 2 - A student's VPython program that models the motion of a fan cart subject to a constant force (constant acceleration/unbalanced forces model). Green boxes highlight where we focus students' attention during model construction.

PhysUtil, for enhancing aspects of performing simulations (e.g., MotionMap in Fig. 1). This software is publicly available.[4,13]

VPython is based on the Python programming language and provides an environment to write simple programs that yield robust three-dimensional simulations (Fig. 1). The VPython programming environment was designed to limit the programmatic statements needed to generate highly visual three-dimensional simulations. Students who receive sufficient computational instruction using VPython are able to successfully model novel situations.[14,15]

Fig. 2 shows sample VPython code that models the motion of a fan cart subject to a single constant force. To construct this model, 9th grade physics students created the objects and assigned their positions and sizes (lines 6 – 7), identified and assigned the other given values and relevant initial conditions (lines 9 – 10 and 12 – 14), calculated the net force acting on the object of interest (line 23), and updated the velocity and position of this object in each time step (lines 24 – 26). This code illustrates the algorithm students are taught to predict the motion of objects given the model for their interactions.[11] The code shown in Fig. 2 produces a highly visual simulation generated from a few program statements. This program represents what students are able to construct after instruction in our 9th grade conceptual physics course.

The program shown in Fig. 2 makes use the PhysUtil module. Developed by a team of Georgia Tech computer science majors, the PhysUtil module was designed to further limit the code-writing needed to create highly visual simulations and to enhance the functionality of VPython to include features of the Modeling Instruction curriculum (e.g., motion maps) without the additional burden of writing complex program statements. At present, we have added four Python classes with PhysUtil: PhysAxis, PhysTimer, MotionMap, and PhysGraph. Each of these classes requires a single initialization line (lines 16 – 19 in Fig. 2), which can be provided to the students, and a single update line in the calculation loop (lines 28, 31, and 33). Detailed documentation on each of these classes and use cases are available online.[13]

To illustrate how our particular brand of numerical computation fits into a typical Modeling Instruction course, we present an activity used in a 9th grade physics course during the second half of the semester. Students employed and extended the parabolic motion model for the motion of an Angry Bird[16] to characterize the motion of a kicked soccer ball.

## A Typical Activity: Modeling a kicked soccer ball

In our modified Modeling Instruction course, we presented projectile motion after students had studied 5 previous models.[7] Students discovered that the constant acceleration model was insufficient to describe the motion of objects in two dimensions subject to the ordinary gravitational force, **F**$_{grav}$ = m**g**. In fact, an appropriate description required the use of two models: the constant acceleration model in the vertical direction and the constant velocity model in the horizontal direction. Typically, the parabolic motion model represents the capstone of the Modeling Instruction curriculum's treatment of force and motion. In our treatment, we used numerical computation to investigate the parabolic motion model, to compare its predictions to real-world observations, and to resolve the limited predictions of this model by extending the model to include air-resistance drag.

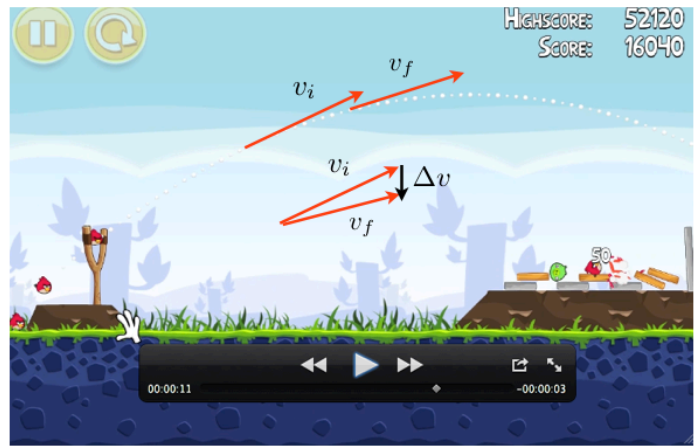

Fig. 3 - A vector construction to determine the direction of the acceleration in the Angry Bird's world. Fair use reproduction (non-profit educational illustration)

Students often collect data from a lab experiment to motivate the development of a new model, but it is also possible to collect data from something that is itself a model (e.g., a computer game). We motivated the parabolic motion model by showing students a snapshot of the trajectory of a bird from the popular Angry Birds video game (Fig. 3). From this vector construction, students concluded that there must be a force acting on the Angry Bird that points vertically downward. To investigate this claim, we collected video data of Angry Birds flying across the screen and then imported this data into Tracker, a free and open-source video tracking software package,[17] where the motion of the Angry Bird was logged and plotted. Tracker allows the user to compute the velocity and acceleration of the tracked particle in each coordinate's direction and to plot those quantities. From their analysis, students determined how to compose the parabolic motion model.

In our course, students also generalized the constant-acceleration computational model that they had developed to the parabolic motion model. Students had previously developed a fully "vector compliant" program to model constant acceleration in either the $x$ or $y$ direction (e.g., lines 23 – 24 in Fig. 2). This generalization emphasizes the interconnected nature of different force and motion models in the Modeling Instruction curriculum. The generalization is quite simple because, in a computational model, the change between different types of motion under constant acceleration is simply a change of the initial conditions. By giving the object an initial velocity with a nonzero horizontal and vertical components, the student can very easily move from modeling an object dropped from a known height (for example) to an object fired into the air at a known angle.

The new computational model linked the different representations of the physical system and provided instant visual feedback about students' physical model. Using PhysAxis, PhysTimer, and PhysGraph, students reconstructed the Angry Bird's motion. Their computational model allowed students to immediately observe if their physical model had any inconsistencies (e.g., unexpected motion in the horizontal direction) or if their computational model had any unrealistic effects (e.g., motion not terminating at the ground level). The latter led to a nice discussion of the limitations of computational models; they can only do what you have told them to do. Using MotionMap, students constructed an animated motion map to observe how components of the force or velocity change with time. With a computational model, students were able to systematically adjust parameters (e.g., the Angry Bird's mass, size, and initial velocity) to observe their effects on the animation; students paid particular attention to their graphs of kinematic and dynamic quantities and their motion maps. Students reported their observations to their peers.

Students were then confronted with the following challenge: "We have learned that the constant acceleration model can help us describe how an object moves in one dimension, and that the parabolic motion model can help us in two dimensions. What about a soccer ball that you kick into the air? How can we model this situation?" Typically, this would be dealt with using the parabolic motion model. By using the computational modeling, we can push this further. "What about a real effects of the air? Do any of these models still apply to the motion?"

Students concluded from video analysis of a kicked ball (similar to the Angry Birds analysis) that there were accelerations in both the horizontal and vertical directions. Moreover, they observed that these accelerations changed with time. Students proposed air resistance as the culprit for this change. However, the model for air resistance (even linear drag) does not lend itself to analytical solutions achievable by 9th grade students; the mathematics is too sophisticated. Computational modeling allowed us to insert a velocity dependent drag force on the ball, and then to use the model to accurately predict the trajectory and landing point for the soccer ball. The motion of real projectiles is no longer intractable to conceptual physics students.

Using numerical computation in the way we described does not supplant the typical activities in which students engage; it enhances and extends those activities.[15] We are making the activities more relevant to students by including real-world examples, emphasizing the concept of models, illustrating the generality of physical principles, and providing a platform for future learning in numerical computation.

### Reflections

We have used numerical computation in a 9th-grade Modeling-Instruction-based Honors Physics course in a private school setting for the last two years, each year comprising a different set of 15-18 students. In that time, we have observed several challenges to student learning and broader adoption.

Students find debugging their programs difficult; that is, they have trouble determining whether they have made a coding error or a physics error and how to deal with that issue. This is likely due to the somewhat loose integration of computational modeling in their physics course. Presently, the length of time between exposures to VPython is too long, and students spend too much time relearning old programming skills. The course requires tighter integration of computational modeling into each assignment and modeling cycle. We have begun providing scaffolded code and performing live coding exercises, both of which are best practices from computer science education. Additionally, we have started to develop our own studies of student thinking and practices (Refs. 14 and 15) to improve instruction.

Resources for computational instruction are not widespread; most materials were developed by Georgia Tech's Physics Education Research group in conjunction with the classroom teacher. However, a virtual community has begun building resources for math and science teachers interested in introducing students to numerical computation. Many of these computational thinking resources are available online.[18] Not all of these resources are tied to the Modeling Instruction curriculum nor are most resources physics related, but the support of such a community could produce additional high-quality resources and can provide support for early adopters, interested teachers, and, most importantly, our students.

## Acknowledgements

The authors would like to thank the students and teachers who have provided constructive feedback on this project over the last few years. Particular thanks to Georgia Tech graduates Cory Johnson, Sebastian Marulanda, Raschel Mead, and Colin Schoeneman for their work to develop PhysUtil .